\documentclass[fleqn,12pt,twoside]{article}
\usepackage{espcrc1}
\usepackage{graphicx}
\usepackage[figuresright]{rotating}

\newcommand{\AmS}{{\protect\the\textfont2
  A\kern-.1667em\lower.5ex\hbox{M}\kern-.125emS}}
\title{Mass effects in hard exclusive photoproduction of $J/\Psi$
mesons}

\author{B. J\"ager\address{Institut f\"ur Theoretische Physik,
Universit\"at Regensburg, \\ D-93040 Regensburg, Germany} and
        W. Schweiger\address{Institut f\"ur Theoretische Physik,
        Universit\"at Graz, Universit\"atsplatz 5 \\ A-8010 Graz, Austria}
	\thanks{Talk given by W. Schweiger at \lq\lq QCD-N02'', Ferrara, Italy, 
	April 2002.}}

\begin{document}


\maketitle

\begin{abstract}
We report on an attempt to describe hard exclusive photoproduction of
$J/\Psi$ mesons, i.e. the reaction $\gamma p \rightarrow J/\Psi p$, by
means of a modified version of the hard-scattering approach, in which
the proton is treated as a quark-diquark rather than a three-quark
system.  In order to improve the applicability of the model at
momentum transfers of only a few GeV we take into account constituent-mass
effects in the calculation of the perturbative scattering amplitude.
With a standard $J/\Psi$-meson distribution amplitude and
diquark-model parameters adopted from preceding investigations of
other photon-induced reactions our predictions for differential cross
sections overestimate the naive extrapolation of the low-momentum
transfer ZEUS data.  Our results, however, reveal the importance of
taking into account the charm-quark mass.
\end{abstract}

\section{INTRODUCTION AND PRESENTATION OF THE MODEL}

The existence of a large momentum or mass scale is the basis of any
perturbative treatment of hadronic processes.  In photoproduction of
heavy quarkonia there are two obvious scales: on the one hand the
heavy quark mass and, on the other hand, the (transverse) momentum
transfer.  If one considers $J/\Psi$ production the charm-quark mass
is still too small to really serve as a hard scale.  It is therefore
likely that significant contributions come from non-perturbative
regions in which hadronic fluctuations of the photon have a large
transverse size.  This statement is supported by the success of
Pomeron phenomenology in describing diffractive photoproduction of
$J/\Psi$s~\cite{DoLa00}.

\begin{figure}[htb]
\begin{minipage}[t]{78mm}
\includegraphics*[width=78mm]{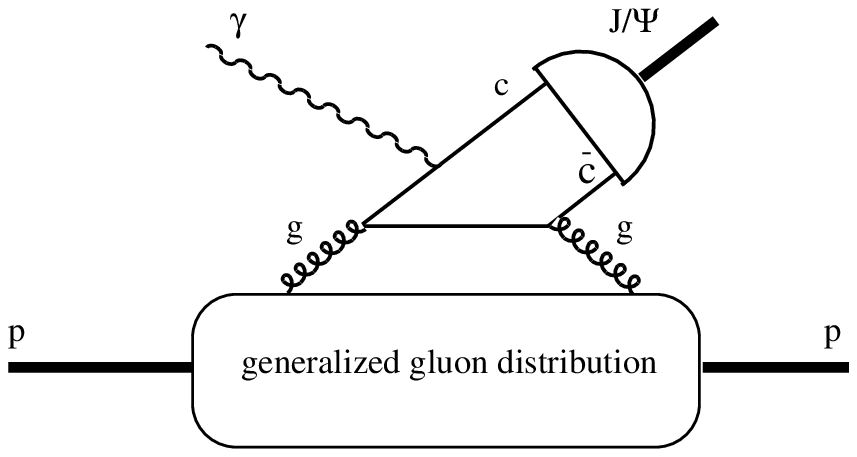} 
\vspace{-1.0cm}
\caption{Handbag-type contribution to $\gamma p \rightarrow J/\Psi p$.}
\label{overlap}
\end{minipage}
\hspace{\fill}
\begin{minipage}[t]{78mm}
\includegraphics*[width=78mm]{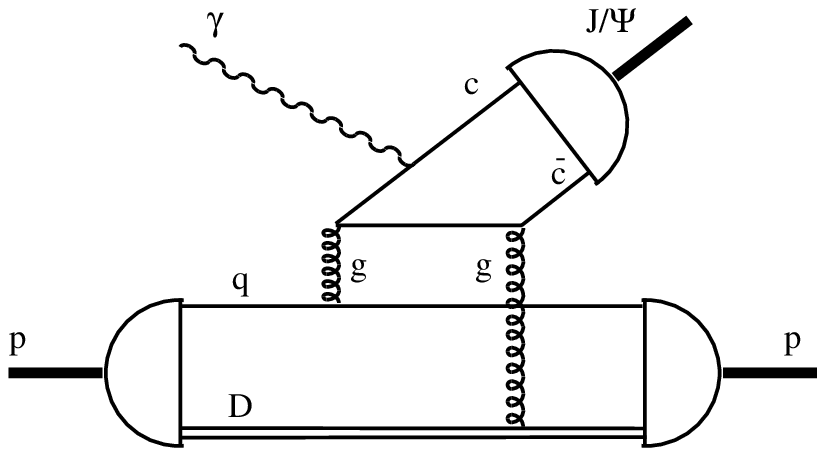} 
\vspace{-1.0cm}
\caption{Diagram representing a typical hard-scattering contribution
for $\gamma p \rightarrow J/\Psi p$ within the diquark model.}
\label{hsm}
\end{minipage}
\end{figure}

Taking, on the other hand, the momentum transfer as a large scale one
can think of several perturbative mechanisms.  A potential
mechanism at intermediate momentum transfers is depicted in
Fig.~\ref{overlap}.  Only one protonic constituent, a gluon,
participates in the hard-scattering subprocess.  A generalized gluon
distribution describes the emission and reabsorption of the gluon.
Such a kind of mechanism has been considered by Huang and Kroll for
the photoproduction of light mesons~\cite{HuKr00}.  They, however,
found a strong suppression of this production mechanism and concluded
that vector-meson dominance is still at work at moderately large
momentum transfers where data exist.

For asymptotically large momentum transfers it is generally assumed
that the hard-scattering mechanism dominates~\cite{LeBr80}.  The
hard-scattering mechanism involves only valence Fock states of the
hadrons with the (leading-order) perturbative part being a coherent
sum of tree graphs specific to the reaction under consideration.  The
non-perturbative ingredients of the hard-scattering mechanism are
hadron distribution amplitudes, i.e. probability amplitudes for
finding the hadronic constituents carrying certain fractions of their
parent hadron's momentum.  This is the reaction mechanism we
concentrate on in our investigation of $J/\Psi$ photoproduction.  As
the graph in Fig.~\ref{hsm} shows, we assume in addition that the
valence Fock state of the proton is a quark-diquark rather than a
three-quark state.  This does not only simplify calculations, but
diquarks also serve to model non-perturbative effects so that the
hard-scattering mechanism may become applicable at reasonably large
momentum transfers where data are available or can be expected.  We
take into account scalar and vector diquarks and treat them as
extended particles with form factors at the gauge-boson diquark
vertices.  For the present analysis the form-factor parameters and
also the quark-diquark distribution amplitude of the proton are
adopted from previous work on baryon form factors, Compton scattering
and other reactions (see, e.g., Ref.~\cite{Ja93}).  The distribution
amplitude for the $J/\Psi$, i.e. $\phi_{J/\Psi}(x) \propto x (1-x)
\hbox{exp}[-b^2 m_{J/\Psi}^2 (x- 1/2)^2]$, is taken from
Ref.~\cite{FeKr97}.  Its normalization is fixed by the experimentally
known $J/\Psi$ decay constant.  These are almost all the ingredients
necessary for calculating the $J/\Psi$-photoproduction amplitude.

\section{TREATMENT OF THE CHARM-QUARK MASS}
At this point it is necessary to mention that constituent masses are
usually neglected in the calculation of the perturbative part of the
amplitude as long as they do not fix the energy or momentum-transfer
scale (like in quarkonium decays).  If we want to make predictions for
momentum transfers which may become experimentally accessible in the
near future, i.e. values not much larger than the charm-quark mass
$m_{c}$, we have to think about taking into account $m_{c}$ in the
perturbative amplitude.  Following a common practice we assume that
the 4-momentum of every hadronic constituent is proportional to the
4-momentum of its parent hadron.  This implies that every constituent
of hadron $H$ has an effective mass $x m_{H}$, with $x$ being the
fraction of the hadron's 4-momentum carried by the constituent. 
Keeping in mind that the perturbative scattering amplitude for the
constituents is convoluted with the hadron distribution amplitudes
this implies, e.g., that $m_{c} \approx m_{J/\Psi}/2$ for our $J/\Psi$
distribution amplitude, which is strongly peaked at $x=1/2$.  This
prescription for constituent masses applies to (on-shell) particles
which show up as external legs in the Feynman diagrams.  To obtain the
constituent mass for an internal line we express the corresponding
momentum as a sum of hadron momenta multiplied with the appropriate
momentum fractions and replace the hadron momenta by their respective
mass.  If the resulting constituent mass contains a term $\propto (x -
y) m_{p}$ this is neglected since it turns out that such terms violate
$U(1)$ gauge invariance.  For our $J/\Psi$ distribution amplitude the
masses in the the charm-quark propagators become in this way again
$m_{c} \approx m_{J/\Psi}/2$.  The Feynman diagrams are then
calculated with these constituent masses and expressed in terms of the
Mandelstam variables $\tilde{s}$, $\tilde{t}$, and $\tilde{u}$
($\tilde{s}+\tilde{t}+\tilde{u} = m_{J/\Psi}^2$).  Keeping the
scattering angle fixed, we finally expand the perturbative amplitude
in terms of $(m_{p}/\sqrt{\tilde{s}})$ and
$(m_{p}/\sqrt{\tilde{s}-m_{{J/\Psi}}^2})$ and neglect proton-mass
corrections of order $(m_{p}^2/\tilde{s})$,
$(m_{p}^2/(\tilde{s}-m_{{J/\Psi}}^2))$, or higher.  This kind of mass
treatment has the nice feature that it preserves $U(1)$ gauge
invariance with respect to the photon and $SU(3)$ gauge invariance
with respect to the gluon.  A similar kind of mass treatment has
already been suggested in Ref.~\cite{BeSch00} for photoproduction of
$\Phi$ mesons.  In that paper, however, massless Mandelstam variables
$\hat{s}$, $\hat{t}$, and $\hat{u}$ have been used instead of
$\tilde{s}$, $\tilde{t}$, and $\tilde{u}$ and the final mass expansion
has not only been applied to the proton but also to the meson.

\section{RESULTS AND CONCLUSIONS}

\begin{figure}[htb]
\begin{minipage}[t]{76mm}
\includegraphics*[width=76mm]{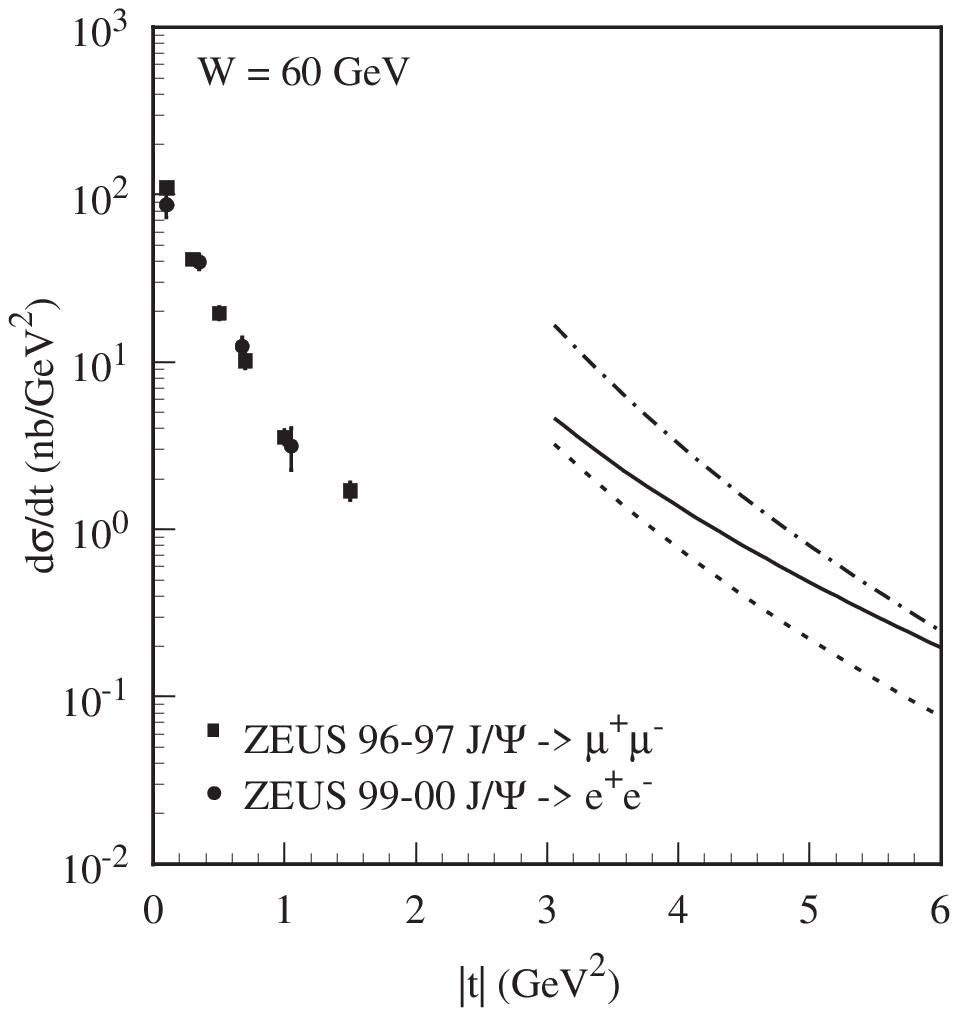} 
\vspace{-1.0cm}
\caption{Diquark-model predictions for the differential cross section
$d\sigma_{\gamma p \rightarrow J/\Psi p}/dt$ at $W=\sqrt{s}=60$~GeV
-- a comparison of different contributions: full calculation,
$\lambda_{J/\Psi}=0,\pm 1$ (solid line), $\lambda_{J/\Psi}=0$ only
(dashed line), mass effects neglected (dash-dotted line). Data are
taken from Ref.~\cite{Che02}.}
\label{jpsi60}
\end{minipage}
\hspace{\fill}
\begin{minipage}[t]{76mm}
\includegraphics*[width=76mm]{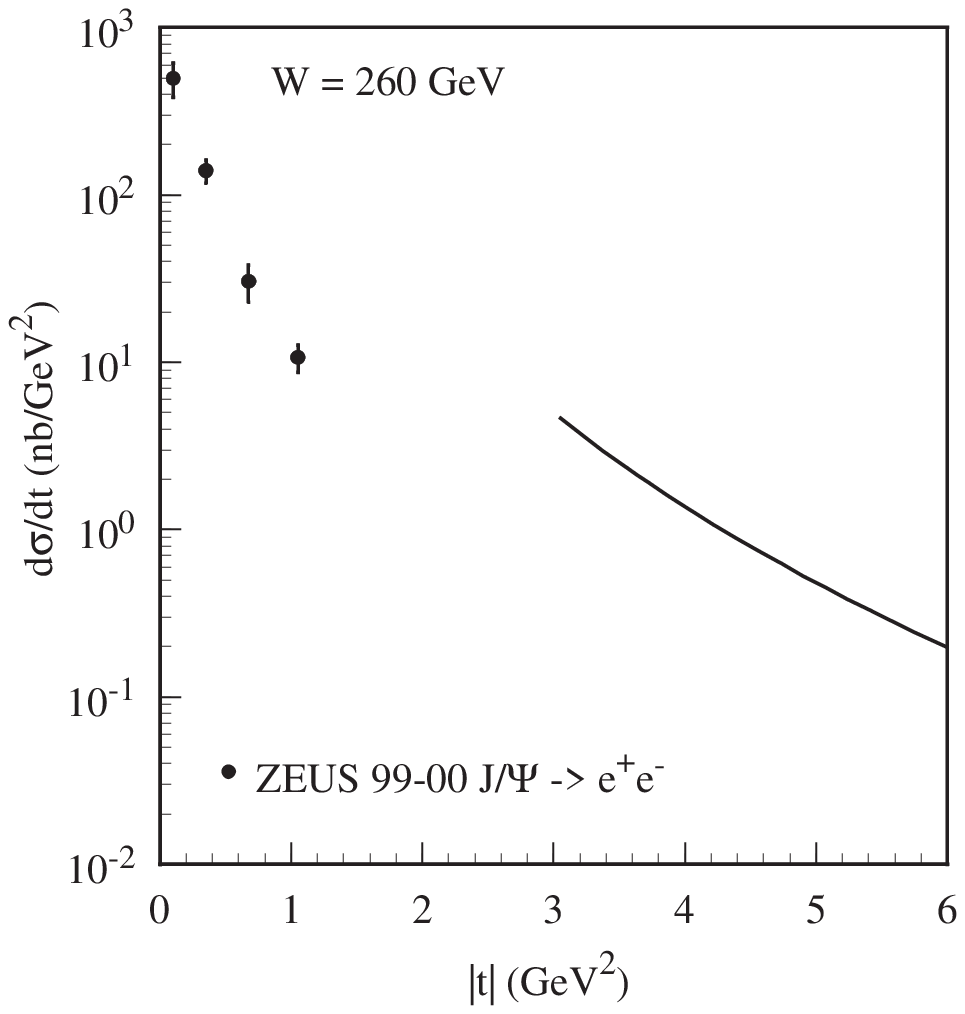} 
\vspace{-1.0cm}
\caption{Diquark-model predictions (full calculation) for the
differential cross section $d\sigma_{\gamma p \rightarrow J/\Psi
p}/dt$ at $W=\sqrt{s}=260$~GeV. Data are taken from
Ref.~\cite{Che02}.}
\label{jpsi260}
\end{minipage}
\end{figure}

Figures~\ref{jpsi60} and \ref{jpsi260} show the diquark-model
predictions for $d\sigma/dt[\gamma p \rightarrow J/\Psi p]$ for two
HERA energies along with recent ZEUS data.  These data are only
available at low momentum transfers ($|t| \leq 1.5$~GeV$^2$), i.e.
outside the validity range of the perturbative diquark model, so that
a direct comparison with our predictions is not possible.  Compared
with a naive extrapolation of the data our predictions, however, seem
to overshoot the $J/\Psi$ photoproduction cross section in the few-GeV
momentum transfer region.  The finite constituent masses provide two
effects which are both sizable, but which partly compensate each
other.  For vanishing constituent masses only two of the twelve
independent helicity amplitudes are different from zero, namely those
for $\lambda_{J/\Psi}=0$ and $\lambda_{p}^{{in}}=\lambda_{p}^{{out}}$
(dash-dotted line in Fig.~\ref{jpsi60}).  Introducing constituent
masses as described above and still taking only these two helicity
amplitudes results in a considerable reduction of the cross section
(dashed line versus dash-dotted line in Fig.~\ref{jpsi60}).  With
finite constituent masses, however, also the other helicity amplitudes
become different from zero which increases the cross section again
(solid line in Fig.~\ref{jpsi60}).  The overall effect of taking into
account constituent masses is thus mainly a flattening of the cross
section towards smaller momentum transfers.  A further flattening and
reduction of the cross section can be expected if one takes into
account that the minimum momentum transfer $t_{min}$ in exclusive
$J/\Psi$ photoproduction is $\approx -1.7$~GeV$^2$.  In order to come
to definite conclusions about the validity of the diquark model for
$J/\Psi$ photoproduction the corresponding modifications for the gluon
kinematics certainly have to be investigated.  Nevertheless, it is
interesting to notice that, in contrast to other exclusive reactions
in which the leading-order perturbative calculation often
underestimates the data, it rather seems to come out too large for
$J/\Psi$ photoproduction.  This, at least, hints at the possibility
that the perturbative production mechanism could provide a sizable
contribution in exclusive $J/\Psi$ photoproduction already at a
few-GeV of momentum transfer.  It would therefore be interesting to
have not only predictions from an effective model like ours, but also
perturbative predictions on the more fundamental pure quark level. 
And finally it would, of course, also be highly desirable to get
experimental cross-section data in the few-GeV momentum-transfer
region to compare with.

%
%


\begin{thebibliography}{9}
\bibitem{DoLa00} A. Donnachie and P.V. Landshoff, Phys. Lett. B 478
(2000) 146.
\bibitem{HuKr00} H. W. Huang and P. Kroll, Eur. Phys. J. C 17 (2000) 423.
\bibitem{LeBr80} G. P. Lepage and S. J. Brodsky, Phys. Rev. D 22
(1980) 2157.
\bibitem{Ja93} R. Jakob et al., Z. Phys. A 347 (1993) 109.
\bibitem{FeKr97} T. Feldmann and P. Kroll, Phys. Lett. 413 (1997) 410.
\bibitem{BeSch00} C. F. Berger and W. Schweiger, Phys. Rev. D 61 (2000)
114026.
\bibitem{Che02} ZEUS Collab., S.~Chekanov et al., preprint
hep-ex/0201043.
\end{thebibliography}
\end{document}